\documentclass[final,5p,twocolumn]{elsarticle}

\usepackage[english]{babel}
\usepackage[utf8]{inputenc}
\usepackage[T1]{fontenc}

\usepackage{amssymb}
\usepackage{amsmath}
\usepackage{multirow}
\usepackage{tabularx}
\usepackage{xcolor}

\usepackage{stfloats}

\usepackage{fancyhdr}
\usepackage[alph]{parnotes}
\makeatletter
\def\parnoteclear{%
    \gdef\PN@text{}%
    \parnotereset
}
\makeatother

\journal{Physics Letters B}

\begin{document}

\begin{frontmatter}
  \title{E2 Rotational Invariants of $0_1^+$ and $2_1^+$ states for $^{106}$Cd: the Emergence of Collective Rotation}
  \author[ornl]{T.~J.~Gray}
  \ead{graytj@ornl.gov}
  \address[ornl]{Physics Division, Oak Ridge National
  Laboratory, Oak Ridge ,37831 ,Tennessee,USA}    
  \author[ornl]{J.~M.~Allmond}
  \author[unc,tunl]{R.~V.~F.~Janssens}
  \address[unc]{Department of Physics and Astronomy, University of North Carolina at Chapel Hill, Chapel Hill, 27599, North Carolina, USA}
  \address[tunl]{Triangle Universities Nuclear Laboratory, Duke University, Durham, 27708, North Carolina, USA}
  \author[irfu]{W.~Korten}
    \address[irfu]{IRFU, CEA, Universit{\'e} Paris-Saclay, Gif-sur-Yvette, F-91191, France}
    \author[anu]{A.~E.~Stuchbery}
 \address[anu]{Department of Nuclear Physics and Accelerator
 Applications, Research School of Physics, Australian National
 University, Canberra, 2601, ACT, Australia}   
  \author[git]{J.~L.~Wood}
  \address[git]{School of Physics, Gerogia Institute of Technology, Atlanta, 30332-0430, Georgia, USA}
  \author[unc,tunl]{A.~D.~Ayangeakaa}
  \author[anl,mil]{S. Bottoni}
    \address[anl]{Physics Division, Argonne National Laboratory, Chicago, 60439, Illinois, USA}
     \address[mil]{INFN Sezione di Milano, Milano, 20133, Itlay}
  \author[llnl,inl]{B.~M.~Bucher}
     \address[llnl]{Lawrence Livermore National Laboratory, Livermore, 94550, California, USA}
  \address[inl]{Nuclear Nonproliferation Division, Idaho National Laboratory, Idaho Falls, 83415, Idaho, USA}
  \author[lbnl]{C.~M.~Campbell}
    \address[lbnl]{Nuclear Science Division, Lawrence Berkeley
      National Laboratory, Berkeley, 94720, California, USA}
  \author[anl]{M.~P.~Carpenter}
  \author[lbnl]{H.~L.~Crawford}
  \author[anl]{H.~David}
  \author[irfu,sur]{D.~Doherty}
      \address[sur]{Department of Physics, University of
      Surrey, Guildford, GU2 7XH, United Kingdom}
  \author[lbnl]{P.~Fallon}
  \author[ornl]{M.~T.~Febbraro}
  \author[ornl]{A.~Galindo-Uribarri}
  \author[ornl]{C.~J.~Gross}
  \author[war]{M.~Komorowska}
  \address[war]{Heavy Ion Laboratory, University of Warsaw, Warsaw, Poland}
  \author[anl]{F.~G.~Kondev}
  \author[anl]{T.~Lauritsen}
  \author[lbnl,ornl]{A.~O.~Macchiavelli}
  \author[war]{P.~Napiorkowsi}
  \author[unam]{E.~Padilla-Rodal}
  \address[unam]{Instituto de Cienias Nucleares, UNAM, AP 70-543, 04510
  Mexico, D.~F., Mexico}   
  \author[ornl]{S.~D.~Pain}
  \author[wu,anl]{W.~Reviol}
    \address[wu]{Department of Physics, Washington University, St.
    Louis, 63130, Missouri, USA}  
  \author[wu]{D.~G.~Sarantites}
  \author[anl]{G.~Savard}
  \author[anl]{D.~Seweryniak}
  \author[llnl]{C.~Y.~Wu}
  \author[ornl]{C.-H.~Yu}
  \author[anl,bnl]{S.~Zhu\fnmark[1]}
  \address[bnl]{Physics Department, Brookhaven National
  Laboratory, Upton, 11973, New York, USA}
  \fntext[1]{Deceased}
  \begin{abstract}
 The collective structure of $^{106}$Cd is elucidated by multi-step Coulomb excitation of a 3.849~MeV/$A$ beam of $^{106}$Cd on a 1.1~mg/cm$^2$ $^{208}$Pb target using GRETINA-CHICO2 at ATLAS. Fourteen $E2$ matrix elements were obtained. The nucleus $^{106}$Cd is a prime example of emergent collectivity that possesses a simple structure: it is free of complexity caused by shape coexistence and has a small, but collectively active number of valence nucleons. This work follows in a long and currently active quest to answer the fundamental question of the origin of nuclear collectivity and deformation, notably in the cadmium isotopes. The results are discussed in terms of phenomenological models, the shell model, and Kumar-Cline sums of $E2$ matrix elements. The ${\langle 0_2^+ ||E2||2_1^+ \rangle}$ matrix element is determined for the first time, providing a total, converged measure of the electric quadrupole strength, $\langle Q^2 \rangle$, of the first-excited $2_1^+$ level relative to the $0_1^+$ ground state, which does not show an increase as expected of harmonic and anharmonic vibrations. Strong evidence for triaxial shapes in weakly collective nuclei is indicated; collective vibrations are excluded. This is contrary to the only other cadmium result of this kind in $^{114}$Cd by C. Fahlander \textit{et al.}, Nucl. Phys. \textbf{A485}, 327 (1988), which is complicated by low-lying shape coexistence near midshell.

 \end{abstract}

  
\end{frontmatter}

\par%
Nuclear collectivity has traditionally been thought to evolve from magic or semi-magic closed-shell nuclei with seniority (pairing) character, through weakly deformed open-shell nuclei with vibrational character, and finally to well-deformed open-shell nuclei with rotational character. This perspective has been the paradigm since the first half of the 20th century, but a series of recent detailed spectroscopy experiments on the cadmium isotopes has challenged this view~\cite{Garret2019, Garret2007, Garret2008, Garret2020}. 
\par%
The stable even cadmium isotopes have been the focus of considerable interest in characterizing the emergence of collectivity, and they have often been considered to be some of the best examples of low-energy vibrators~\cite{Rowe1970, Casten1990, Heyde2004}. The energy levels of the nearly degenerate multiplet of $0^+$, $2^+$, and $4^+$ states observed in Cd isotopes near midshell are in good agreement with the predictions of a 2-phonon vibrational multiplet. More specifically, the midshell nuclei $^{110,112}$Cd were proposed as vibrational candidates for the U(5) limit of the Interacting Boson Model (IBM) by Arima and Iachello~\cite{Arima1976}; $^{118}$Cd was highlighted as an example of near-harmonic quadrupole vibration~\cite{Aprahamian1987}; $^{110}$Cd was suggested as a good  anharmonic quadrupole vibrator~\cite{Caprio2002}; and, generally, the Cd isotopes have been cited as examples of spherical vibrators by many others (see, e.g.~\cite{Kern1995}). However, two-proton transfer reactions strongly populate the first-excited $0^+$ states in $^{110,112}$Cd, indicating that these levels are intruder bandheads with 2p-4h character, rather than 2-phonon vibrational excitations~\cite{Fielding1977}. The predominant view for some time was that the intruder and 2-phonon $0^+$ states in $^{110,112}$Cd are strongly mixed; such would explain both the decay patterns and transfer data~\cite{Heyde1982,Fortune1987}. In recent years, however, this view has been challenged by high-precision lifetime and branching ratio measurements~\cite{Garret2007,Garret2008}. Multiple shape coexistence in $^{110,112}$Cd was proposed in Refs.~\cite{Garret2019,Garret2020}, where the low-lying level schemes were explained without reference to nuclear vibrations.
\par%
The nucleus $^{106}$Cd ($Z=48, N=58$) is an excellent laboratory for studying the emergence of collectivity. It has 2 valence proton holes and 8 valence neutrons outside the double-magic $^{100}$Sn core, and is the lightest (most proton-rich) stable cadmium isotope. Detailed spectroscopy is available and the $2_2^{+}$, $0_2^{+}$, and $0_3^{+}$ states are known. This nucleus is also sufficiently close to $^{100}$Sn that large-scale shell-model calculations can be applied and used to investigate the interplay between collective and single-particle degrees of freedom. Additionally, the $0_2^+$ and $0_3^+$ states (intruder / shape-coexistence candidates) are lowest in energy at midshell, potentially interfering with the other structural features. For $^{106}$Cd, these states are higher in energy, meaning that a test for low-lying vibrations can be conducted with fewer complications from intruder mixing. 

\par%
There has been some interest in the light Cd isotopes, specifically $^{106}$Cd, in recent years~\cite{BenczerKoller2016, Schmidt2017, Zhong2020, Rhodes2021, Siciliano2021, Zuker2021, Sharma2021}. The first $B(E2)$ measurements for $^{106}$Cd are from safe Coulomb excitation using Cd targets and proton, $\alpha$, and $^{16}$O beams~\cite{Stelson1958, Milner1969,Esat1976}. However, in 2016 a lifetime measurement using a Doppler-broadened lineshape analysis yielded $\tau(2_1^{+})$ and $\tau(4_1^{+}$) lifetime values in disagreement with the previous Coulomb excitation results~\cite{BenczerKoller2016}. Since then, a multi-step Coulomb excitation experiment using $^{208}$Pb and $^{48}$Ti targets has been published~\cite{Rhodes2021}, confirming the previous Coulomb excitation results, in disagreement with Ref.~\cite{BenczerKoller2016}. Moreover, two more recent lifetime measurements~\cite{Zhong2020,Siciliano2021} (recoil distance Doppler-shift and decay-curve methods, respectively) of $^{106}$Cd also disagree with the results in Ref.~\cite{BenczerKoller2016}.
\par%
Shell-model calculations for $^{106}$Cd using different interactions have been published, with an emphasis on using the shell-model results with Kumar-Cline invariants to predict nuclear shapes~\cite{Schmidt2017, Zuker2021, Rhodes2021}. An alternate approach is taken in Ref.~\cite{Garret2019}, where beyond mean field calculations predict multiple shape coexistence for the $^{110,112}$Cd isotopes. These calculations were extended to $^{106}$Cd in Ref.~\cite{Siciliano2021}, again predicting multiple shape coexistence. Beyond mean field calculations were also used to compare electromagnetic properties of the yrast states in Ref.~\cite{Sharma2021}.
\par
\begin{figure}[h]
  \includegraphics[width=\columnwidth]{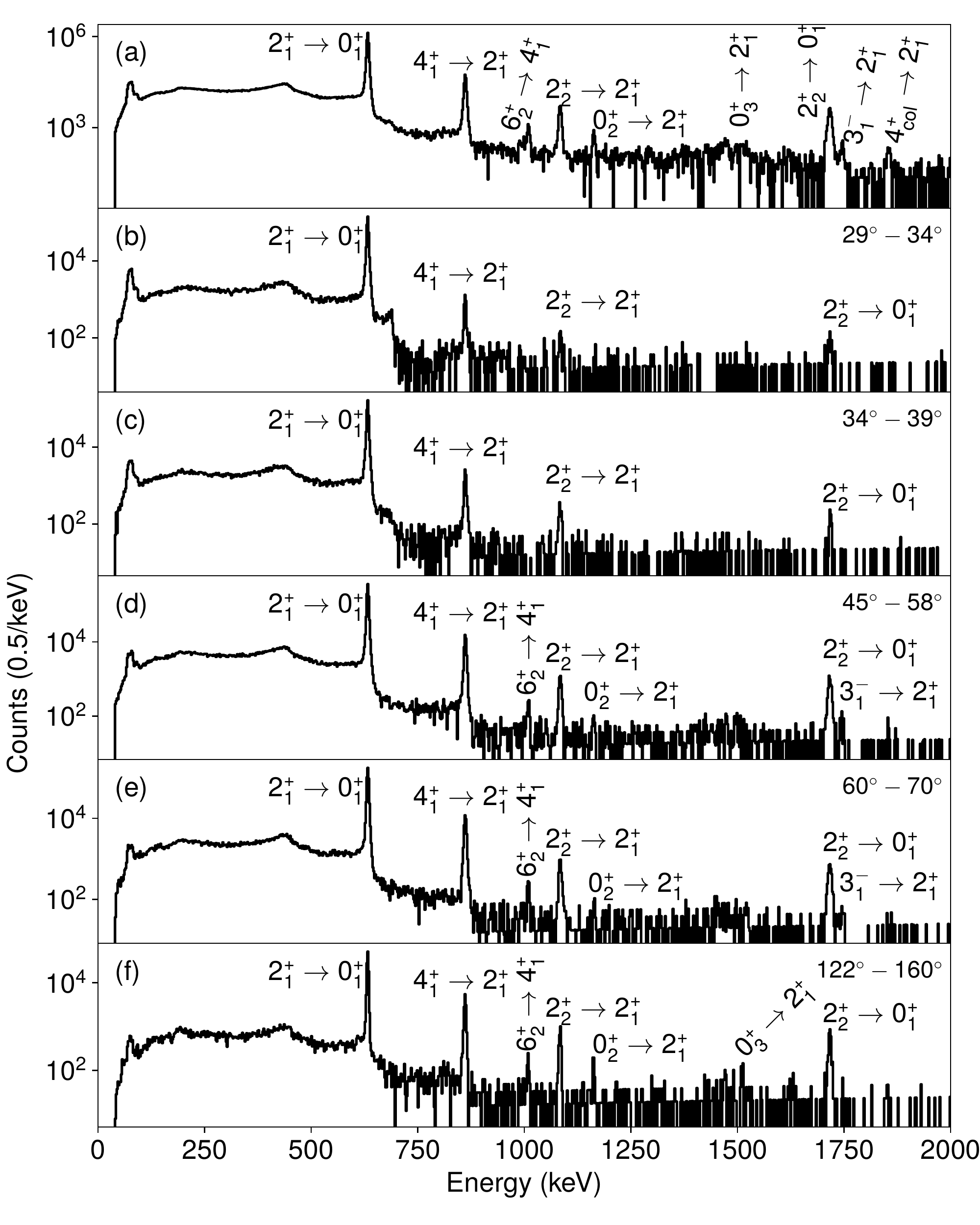}
  \caption{Doppler-corrected $\gamma$-ray spectra for (a): all particle angles, and (b)-(f): the five particle angle regions used in the \textsc{gosia} analysis. The state denoted $4_{\mathrm{col}}$ is the $4^{+}$ state at 2486~keV in $^{106}$Cd.}
  \label{fig:spec_part}
\end{figure}

%
In an effort to help clarify the collective nature of this nucleus, a multi-step Coulomb excitation experiment on $^{106}$Cd was conducted using the ATLAS facility at Argonne National Laboratory. A 408-MeV (3.849 MeV/$A$) $^{106}$Cd beam was incident on a 1.1~mg/cm$^2$, $^{208}$Pb target. A 40-$\mu$g/cm$^2$ thick layer of $^{12}$C was present upstream of the $^{208}$Pb layer. The GRETINA array~\cite{GRETINA} with 32 crystals present was used to detect gamma rays, and the CHICO2 charged particle array~\cite{CHICO,CHICO2} measured recoiling target and beam nuclei. The detected particle angles were used together with an empirical estimate of the target recoil energy to Doppler correct the detected gamma rays. The total gamma-ray spectrum for all particle angles is presented in Fig.~\ref{fig:spec_part} (a). Strong population of the low-lying states --- $2_1^{+}, 4_1^{+}$, $2_2^{+}$ --- is seen, along with weaker population of higher-lying levels --- $6_2^{+}$, $3_1^{-}$, and $4_{\mathrm{col}}^{+}$ at 2486~keV (the designation $4^{+}_{\mathrm{col[lective]}}$ is used as there are an unknown number of $4^{+}$ states at lower energy that are not populated by Coulomb excitation). Weak population of both the $0_2^{+}$ and $0_3^{+}$ states is observed as well.
\par%
The semi-classical Coulomb excitation program \textsc{gosia} was used to extract matrix elements from the gamma-ray intensities~\cite{GOSIA}. The data were separated into five ranges of particle scattering angle. The separate gamma-ray spectra for each of these are given in Figs.~\ref{fig:spec_part} (b)--(f). Experimental input lifetimes ($2_1^{+}$ and $0_2^{+}$), branching ratios, and mixing ratios from Refs.~\cite{Pritychenko2016, Flanagan1976, Daniere1977, Samuelson1979, Kumpulainen1992, Roussiere1984, SchmidtThesis, Linnemann2007, Zhong2020, Siciliano2021} were included as constraints in the \textsc{gosia} $\chi^2$ analysis. Individual constraints were removed to check the self-consistency of the results obtained. 
There is generally little to no sensitivity to the signs of the quadrupole interference terms, which are products of transition matrix elements. There is a slight preference in the chi-squared surfaces for positive $P_3 = \langle 0_1^{+} || E2 || 2_2^{+} \rangle \langle 2_2^{+} || E2 || 2_1^{+} \rangle \langle 2_1^{+} || E2 || 0_1^{+} \rangle$ and both signs give approximately equivalent matrix element magnitudes. A positive $P_3$ term is consistent with the majority of nuclei, including $^{114}$Cd~\cite{Allmond2009, Fahlander1988}.
Systematic uncertainties accounting for energy loss through the target, relative $\gamma$-ray detection efficiency, unknown $\delta(M1/E2)$ mixing ratios, and unknown matrix elements for unobserved transitions, were accounted for. The extracted matrix elements are given in Table~\ref{tab:me}.
\begin{table}
  \centering
  \caption{Matrix elements extracted from this work compared to
    previous results.}
  \label{tab:me}
\begin{tabularx}{\columnwidth} { 
  m{0.3cm}
  m{0.35cm}
  m{0.4cm}
  >{\centering\arraybackslash}p{1.6cm}
  >{\centering\arraybackslash}p{1.5cm}
  >{\centering\arraybackslash}X
}
  \multicolumn{6}{c}{$\langle J_i^{\pi} || E2 || J_f^{\pi}\rangle$ (eb)}                                                 \\ \hline
  $J_i^{\pi}$   & $J_f^{\pi}$ & $E_{\gamma}$ & This work   & Ref.~\cite{Rhodes2021} & Other Refs.                        \\ \hline
  $0_1^+$       & $2_1^+$     & 633          & $0.636(9)$  & $0.652(11)$            & $0.76(3)$~\cite{BenczerKoller2016} \\
                &             &              &             &                        & $0.653(13)$~\cite{Milner1969}      \\
                &             &              &             &                        & $0.620(3)$~\cite{Esat1976}         \\
  $2_1^+$       & $ 4_1^+$    & 861          & $1.05(3)$   & $1.044(24)$            & $0.79(2)$~\cite{BenczerKoller2016} \\
                &             &              &             &                        & $1.11(7)$~\cite{Milner1969}        \\
  $4_1^+$       & $ 6_2^+$    & 1009          & $1.18(9)$   & $1.37(10)$             &                                    \\
  $2_1^+$       & $ 2_2^+$    & 1084         & $0.44(3)$   & $0.415(15)$            & $0.49(4)$~\cite{Milner1969}        \\
                &             &              &             &                        & $0.32(5)$~\cite{Grabowski1973}     \\
  $0_1^+$       & $ 2_2^+$    & 1716         & $0.195(15)$ & $0.169(4)$             & $0.190(13)$~\cite{Milner1969}      \\
  $2_1^+$       & $ 0_2^+$    & 1163         & $0.176(14)$ &                        &                                    \\
  $2_2^+$       & $ 0_2^+$    & 79           & $< -0.02$   &                        &                                    \\
  $2_1^+$       & $ 4_{\mathrm{col}}^+$\parnote[a]{State energy is $2486$~keV.\label{mk:en}}    & 1853         & $0.09(4)$   &                        &                                    \\
  $2_2^+$       & $ 4_{\mathrm{col}}^+$\parnoteref{mk:en}    & 770          & $0.26(12)$  &                        &                                    \\
  $2_2^+$       & $ 0_3^+$    & 427          & $0.4(2)$    &                        &                                    \\
  $2_1^+$       & $ 0_3^+$    & 1511         & $0.026(13)$ &                        &                                    \\ 
  $2_1^+$       & $ 2_1^+$    &              & $-0.38(17)$ & $-0.25(5)$             & $-0.37(11)$~\cite{Esat1976}        \\
  $4_1^+$       & $ 4_1^+$    &              & $-0.15(18)$ & $-0.52(24)$            &                                    \\
  $2_2^+$       & $ 2_2^+$    &              & $+0.81(38)$ & $1.33(6)$              &                                    \\
  \multicolumn{6}{c}{$\langle J_i^{\pi} || M1 || J_f^{\pi}\rangle$ ($\mu_N$)}                                            \\ \hline
  $J_i^{\pi}$   & $J_f^{\pi}$ & $E_{\gamma}$ & This work   & Ref.~\cite{Rhodes2021} & Other Refs.                        \\ \hline
  $2_1^+$       & $ 2_2^+$    & 1084         & $-0.31(3)$  & $-0.263(17)$           & $-0.39$~\cite{Milner1969}          \\
                &             &              &             &                        & $-0.35(5)$~\cite{Grabowski1973}    \\
  \multicolumn{6}{c}{$\langle J_i^{\pi} || E3 || J_f^{\pi}\rangle$ (eb$^{3/2}$)}                                         \\ \hline
    $J_i^{\pi}$ & $J_f^{\pi}$ &              & This work   & Ref.~\cite{Rhodes2021} & Other Refs.                        \\ \hline
  $0_1^+$       & $ 3_1^-$    & 2379         & $<0.4$      & $0.28(14)$             & $0.40(5)$~\cite{Fewell1985}        \\
                                           
\end{tabularx}
\parnotes
\end{table}
\par%
The results compare well with previously published values, and mostly agree with the recent Coulomb excitation results~\cite{Rhodes2021}. It is clear that the $2_1^{+}$ and $4_1^{+}$ lifetime results from Ref.~\cite{BenczerKoller2016} are anomalous and do not agree with any of the Coulomb excitation studies; note that no $\tau(4_1^{+})$ constraints were used in the present study. The lifetime of the $0_2^+$ state was measured in Ref.~\cite{Siciliano2021}, but the lack of a known $0_2^+ \rightarrow 2_2^+$ branch prevented a $B(E2; 0_2^+ \rightarrow 2_2^+)$ determination. We report a $\langle 2_1^+||E2||0_2^+\rangle$ value here by using the lifetime in Ref.~\cite{Siciliano2021} as a constraint to the GOSIA fit of observed intensities and evaluating the sensitivity of the extracted matrix element relative to the $\langle 2_2^+||E2||0_2^+ \rangle$ one.  The $0_3^+ \rightarrow 2_1^+$ transition was very weakly observed, meaning that there is little sensitivity to the $\langle 2_2^+ || E2 || 0_3^+ \rangle$ and $\langle 2_1^+ || E2 || 0_3^+ \rangle$ matrix elements. 
\begin{figure*}[h]
  \centering
  \includegraphics[width=0.8\textwidth]{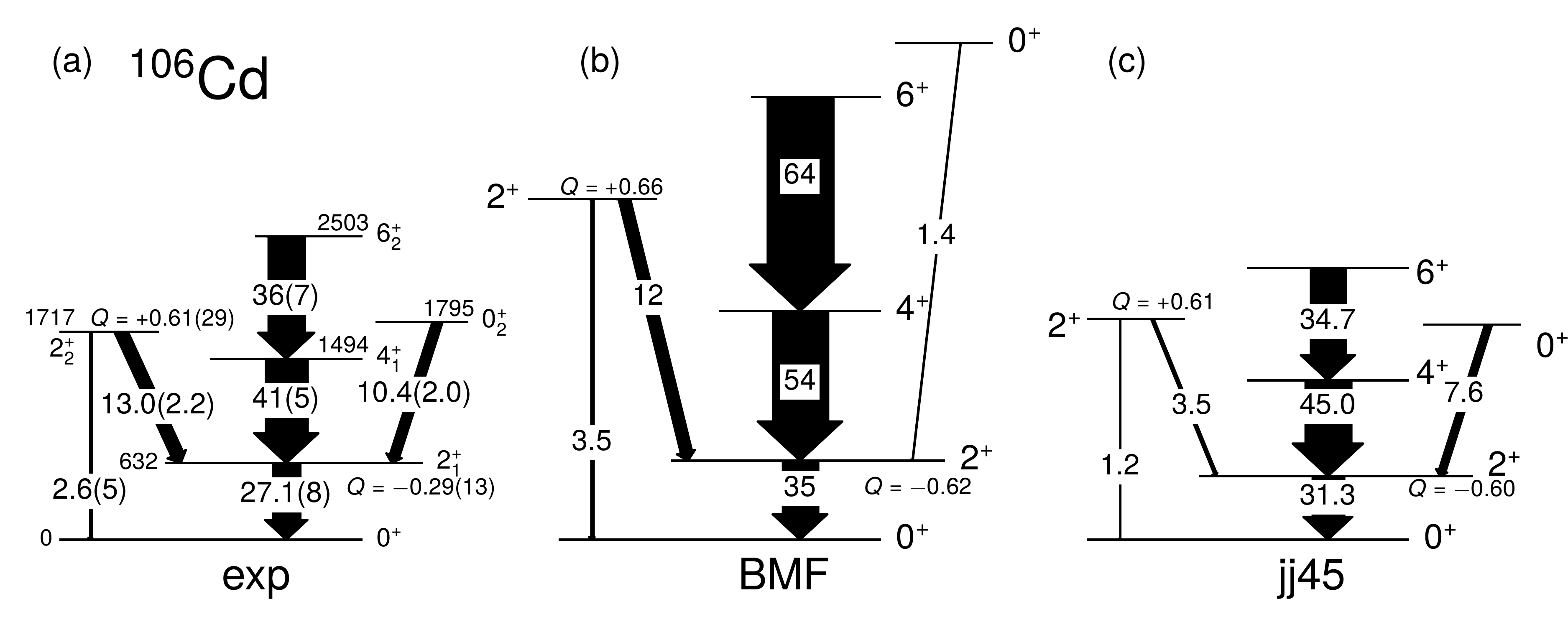}
  \caption{Comparison of low-lying level schemes from experiment, beyond mean-field calculations~\cite{Siciliano2021}, and ``jj45'' shell-model calculations~\cite{Rhodes2021}. Transitions strengths in W.u., $Q_s(2^+)$ values in $e$b, and level energies in keV are indicated.}
  \label{fig:levelschemes}
\end{figure*}
\par%
Electric quadrupole invariants were calculated with the Kumar-Cline sum rules by taking sums over $E2$ matrix elements~\cite{Cline1986}. For the lowest excited states, sufficient experimental information is available to give good confidence that such sums have converged. The two expressions used for the present analysis are:
\begin{align}
  &\langle Q^2 \rangle = \sqrt{\frac{5}{2s + 1}} \sum_r \langle s || E2 || r \rangle \langle r || E2 || s \rangle
                        \begin{Bmatrix} 2 & 2 & 0 \\ s & s & r \end{Bmatrix} \\ \nonumber
  &\langle Q^3 \cos(3\delta) \rangle = (-)^{s+1}\sqrt{\frac{175}{2(2s+1)}} \sum_{rt} (-)^r \\ \nonumber
                      &\qquad \times \langle s || E2 || t \rangle \langle t || E2 || r \rangle \langle r || E2 || s \rangle \\ 
                      &\qquad \times \begin{Bmatrix} 2 & 2 & 2 \\ r & s & t \end{Bmatrix} \begin{Bmatrix} 2 & 2 & 0 \\ s & s & r \end{Bmatrix},                  
\end{align}
where $s$ is the state for which the invariants are evaluated, $r$ and $t$ are intermediate states, and the Wigner 6j symbols are used. The first invariant $\langle Q^2 \rangle$ gives a model independent indication of deformation, while the second one, $\langle Q^3 \cos(3\delta) \rangle$, allows the axial asymmetry $\delta$ to be evaluated with $\delta = \arccos(\langle Q^3 \cos(3\delta) \rangle/\langle Q^2 \rangle^{3/2})/3$. The invariants evaluated for the low-lying states are given in Table~\ref{tab:inv}. While enough matrix elements for the $0_1^+$ and $2_1^+$ states are available to ensure good convergence of the sum, in part due to the new ${\langle 0_2^+ ||E2||2_1^+ \rangle}$ matrix element, only partial sums for the $0_2^+$, $2_2^+$, and $4_1^+$ levels can be evaluated. Thus, lower limits on the $\langle Q^2 \rangle$ invariants are given; matrix elements connecting these states to higher excitations are necessary to evaluate the converged sums. A less likely $P_3<0$ solution has no appreciable impact on the $\langle Q^2 \rangle$ values.

\begin{table}
  \centering
  \caption{Electric quadrupole shape invariants (Kumar-Cline sum rules) extracted from this work.}
  \label{tab:inv}
  \begin{tabularx}{\columnwidth}{
      >{\centering\arraybackslash}X
      >{\centering\arraybackslash}X
      >{\centering\arraybackslash}X
      >{\centering\arraybackslash}X
    }
  \multicolumn{4}{c}{$\langle Q^2 \rangle$ ($e^2$b$^2$)}                \\ \hline          
  State   & Experimental & BMF        & jj45                          \\ \hline
  $0_1^+$ & $0.443(13)$\parnote[a]{A less likely $P_3 < 0$ solution leads to $\langle Q^2 \rangle_{0^{+}} = 0.428(12)$, $\langle Q^2 \rangle_{2^{+}} = 0.373(31)$, $\langle Q^2 \cos(3 \delta) \rangle = +0.20(5)$, and $\langle \delta \rangle = 15(5)^{\circ}$.} \label{mk:p3}  & $0.575$     & $0.484$                                \\
  $2_1^+$ & $0.375(29)$\parnoteref{mk:p3}  & $0.574$   & $0.493$                                 \\
  $4_1^+$ & $>0.28(3)$   & $0.838$     & $0.482$                               \\
  $2_2^+$ & $>0.21(13)$  & $0.661$     & $0.408$                               \\
  $0_2^+$ & $>0.031(5)$  & $1.03$      & $0.0919$                \vspace{0.5em}              \\
  \multicolumn{4}{c}{$\langle Q^3 \cos(3 \delta) \rangle$ ($e^3$b$^3$)} \\ \hline                                     
  State   & Experimental & BMF        & jj45                             \\ \hline
  $0_1^+$ & $+0.01(6)$\parnoteref{mk:p3}   & $+0.202$   & $+0.265$                 \vspace{0.5em}                \\
  \multicolumn{4}{c}{$\langle \delta \rangle$ ($^{\circ})$} \\ \hline                                     
  State   & Experimental & BMF        & jj45                             \\ \hline
  $0_1^+$ & $29(4)$\parnoteref{mk:p3}   & $21$          & $13$                           \\             
\end{tabularx}
\parnotes
\end{table}

\par%
Figure~\ref{fig:levelschemes} (a) shows the low-lying level scheme extracted from this work, with arrow widths corresponding to the $B(E2;\downarrow)$ strengths in Weisskopf units and $Q_s(2^+)$ values in $e$b. The experimental results are compared with two state-of-the-art theoretical approaches: the beyond mean-field calculation (BMF) for $^{106}$Cd in Ref.~\cite{Siciliano2021} (the same as for $^{110,112}$Cd in Refs.~\cite{Garret2019,Garret2020}), and a shell-model calculation presented in Ref.~\cite{Rhodes2021}. The latter is denoted ``jj45'', see Ref.~\cite{Rhodes2021} for details. The results from a second calculation, ``sr88'', were also presented in Ref.~\cite{Rhodes2021}, and are not given here. The ``jj45'' calculation has a more extensive basis than ``sr88'', and for the purposes of the present analysis, including the shape invariants, the two calculations give similar results. The BMF calculations do a good job of reproducing the excitation strengths of the $2_1^+$, $4_1^+$, and $2_2^+$ states. However, the energies predicted by the BMF model are much higher than experimentally observed, almost by a factor of 2. Moreover, the energy ($E_x$) ordering of the excited $0^+$ states is incorrect, and the strength from the $0^+$ level corresponding to the experimental $0_2^+$ state is underpredicted by a factor of $\approx 7$. In contrast, the shell model achieves good agreement with energy levels and predicts the decay strength of the $0_2^+$ well, while underestimating the decay strengths from the $2_2^+$ state. The experimental and theoretical results all show $Q(2_1^+) + Q(2_2^+) \approx 0$, consistent with a rotor, 2-phonon mixing, and global expectations, cf. Fig. 2 in Ref.~\cite{Allmond2013}.
\par%
\begin{figure}[h]
  \includegraphics[width=\columnwidth]{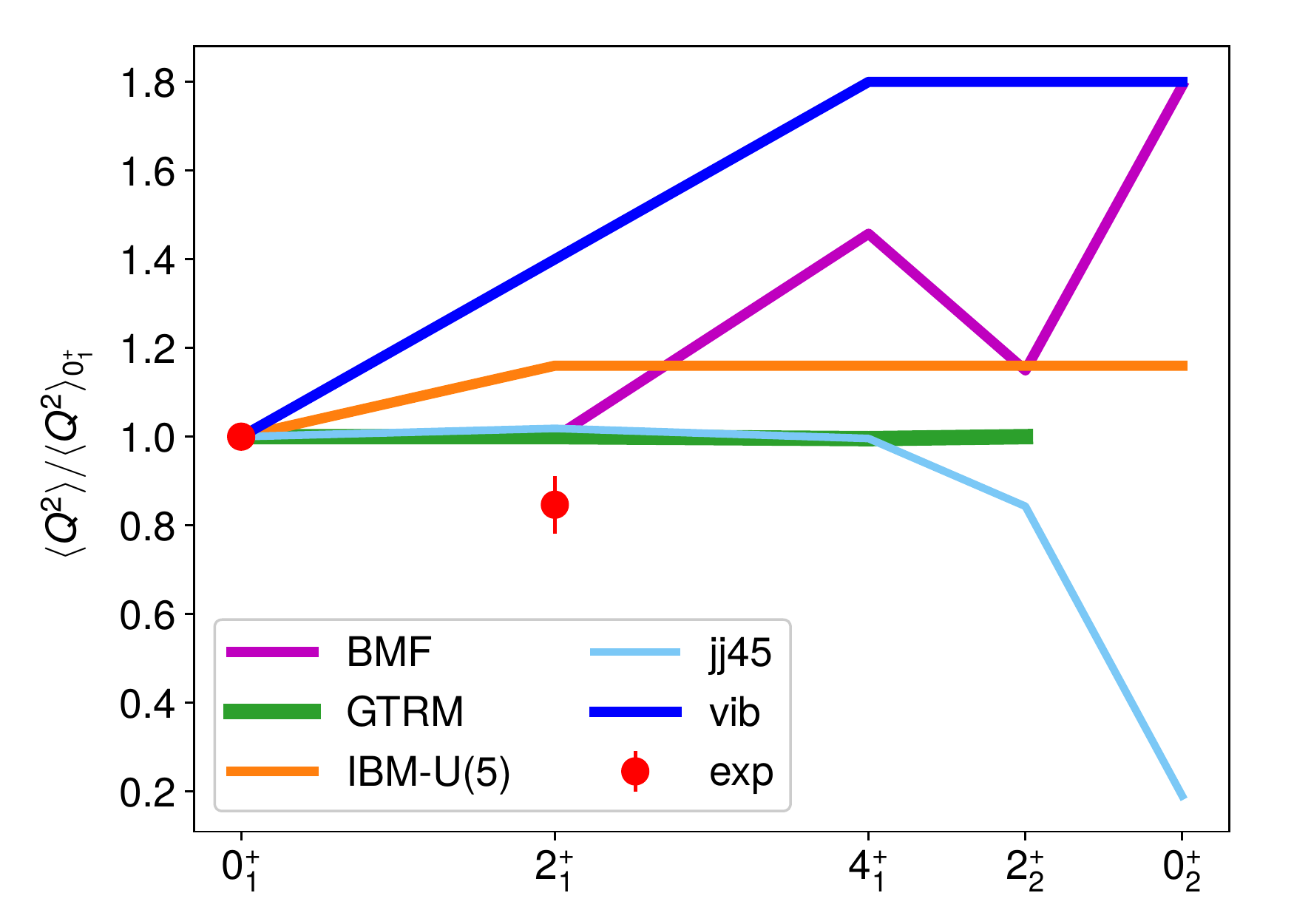}
  \caption{Comparison of electrical quadrupole invariant $\langle Q^2 \rangle$ values for low-lying states between experiment and several theoretical approaches. Geometric vibrations (vib) and vibrations with quenched boson number (IBM-U(5)) are excluded by the experimental $\langle Q^2 \rangle$ value for the $2_1^{+}$ state. Beyond mean-field (BMF), shell-model (jj45), and generalized triaxial rotor model (GTRM) calculations all give equivalent ratios for the $2_1^{+}$ state. $\langle Q^2 \rangle$ sums for the $4_1^{+}, 2_2^{+}$, and $0_2^{+}$ states would differentiate between these models, however more matrix elements are needed for convergence.}
  \label{fig:q2plot}
\end{figure}
\begin{figure*}[h]
  \centering
  \includegraphics[width=0.8\textwidth]{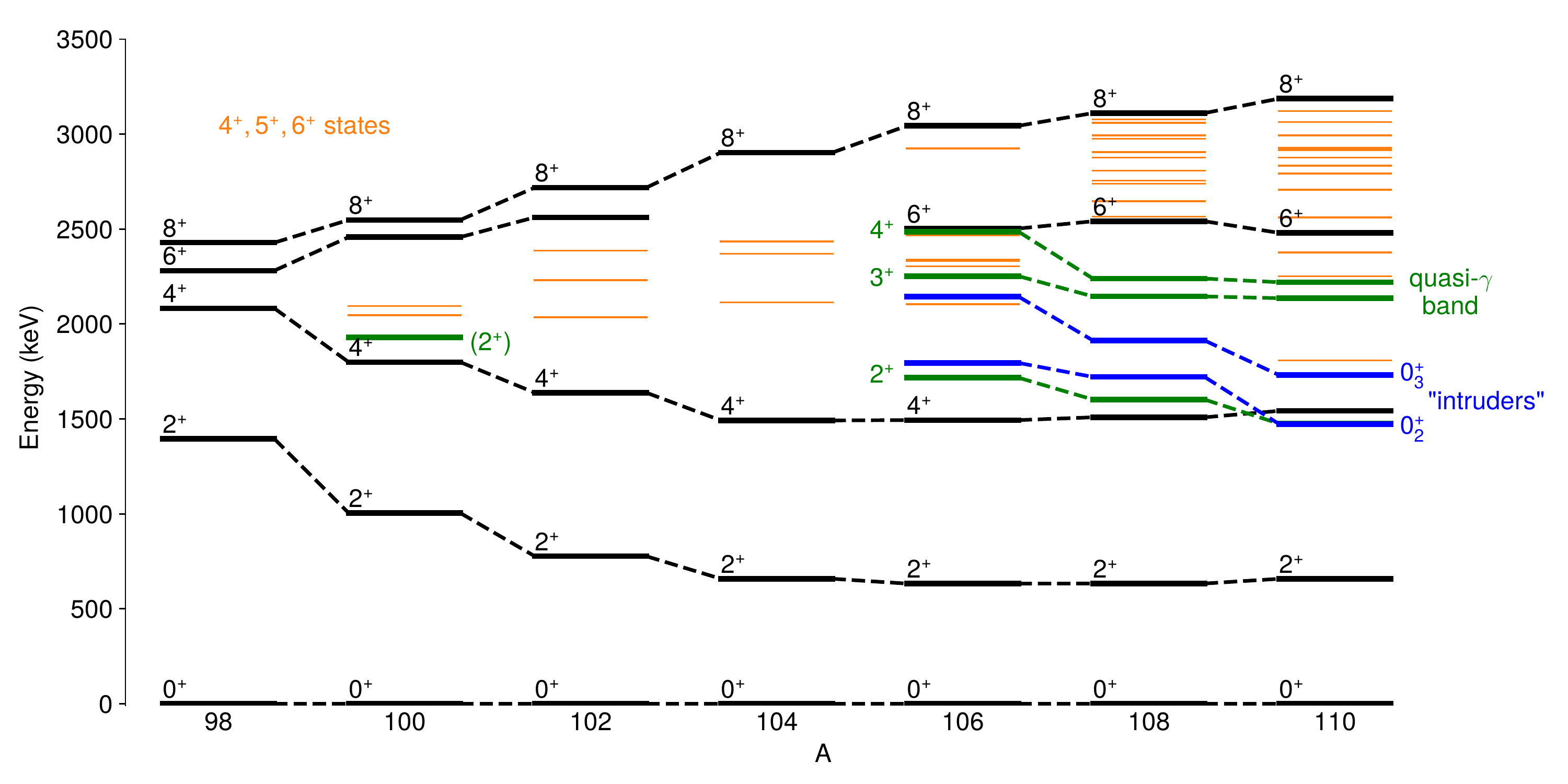}
  \caption{Systematics of the low-mass Cd isotopes. The seniority structure is shown by the black states, with more collective states in blue and green appearing when closer to the mid shell. The green states are a suggested gamma band, and the excited $0^{+}$ states are potentially intruder configurations. The orange states are likely quasi-particle excitations in the valence space.}
  \label{fig:syst}
\end{figure*}
The rotational invariants calculated with the Kumar-Cline sum rules bridge the laboratory and body frames, allowing shape information to be compared in a model independent way. Figure~\ref{fig:q2plot} provides the ratio $\langle Q^2 \rangle / \langle Q^2 \rangle_{0_1^+}$ for low-lying states in $^{106}$Cd. The experimental sums are compared to various theoretical approaches. A geometric vibrator (vib, dark blue) is clearly ruled out by this comparison: the ratio $\langle Q^2 \rangle_{2_1^+} / \langle Q^2 \rangle_{0_1^+} = 1.4$ for a harmonic vibration, 1.31 for a two phonon-mixing treatment of an anharmonic vibration \cite{McGowan1965,Tamura1966, Bes1969, Qin2016}, 1.25 for an effective field theory (EFT) treatment of an anharmonic vibration \cite{Perez2015}, and $1.2-1.4$ for a wide range of anharmonic $\beta$-$\gamma$ potentials within the Geometric Collective Model (GCM) \cite{Troltenier1991}; the experimental value is $0.85(7)$. This is contrary to the only other cadmium result of this kind in $^{114}$Cd by C. Fahlander \textit{et al.} \cite{Fahlander1988}, which showed an increase in $\langle Q^2 \rangle$ but was complicated by low-lying shape coexistence near midshell. An IBM calculation in the $U(5)$ limit was carried out to assess the effect of a finite number of bosons (IBM-U(5), orange). The results of the 5-boson calculation quenched the $\langle Q^2 \rangle$ sum to $1.16$. This is closer to the experimental value than a geometric vibrator, but still not in agreement. The experimental value is close to a triaxial rotor (GRTM, green, $1.0$), the BMF calculations (magenta, $1.0$), and the ``jj45'' SM calculations (light blue, $1.0$). Unfortunately, the experimental sums for the $4_1^+$, $2_2^+$, and $0_2^+$ states are incomplete, and so cannot be meaningfully compared to the theoretical sums. The asymmetry parameter $\delta$ for the ground state is extracted from the first and second quadrupole invariants as $\langle \delta \rangle = 29(4)^{\circ}$, where $0^{\circ}$ is associated with a prolate shape, $30^{\circ}$ with triaxial, and $60^{\circ}$ with an oblate one.
\par%
The results for $^{106}$Cd suggest that the isotopic chain can be described as evolving directly from seniority to rotational character with competition from intruders or shape coexistence becoming progressively more influential towards midshell, in line with the explanation proposed for $^{110,112}$Cd by Garrett et. al~\cite{Garret2019}. However, the BMF calculations do not describe the excited $0^+$ states for $^{106}$Cd as well as they do for $^{110,112}$Cd. Thus, a transition in the character of the $0_2^+$ levels may occur between $^{106}$Cd and $^{110}$Cd or essential physics is missing in the theoretical description. Figure~\ref{fig:syst} displays the energy and systematics across the proton-rich Cd nuclei, starting with semi-magic $^{98}$Cd at the $N=50$ shell closure. The present work on $^{106}$Cd shows that the $4^{+}$ state at 2486~keV and the $6^{+}$ level at 2503 keV are populated by Coulomb excitation while several lower-lying $4^{+}$ states, and a $6^{+}$ one at 2492 keV (Refs.~\cite{Kumpulainen1992,Roussiere1984}), that are not observed. Coulomb excitation selectively populates states connected by strong $E2$ matrix elements --- i.e., it differentiates between collective and non-collective structures. Thus, the current data may suggest that the $6_2^{+}$ level should be associated with the $0_1^+, 2_1^+, 4_1^+$ sequence, and that the $4^{+}$ state at 2486 keV may be connected with a ``$\gamma$ band'' member, associated with the $2_2^{+}$ and $3_1^{+}$ states. 
\par%
In summary, $^{106}$\textrm{Cd} was studied by multi-step Coulomb excitation on a $^{208}$Pb target with GRETINA-CHICO2 and some 14 $E2$ matrix elements were extracted. These matrix elements were compared to several theoretical approaches. The rotational invariants $\langle Q^2 \rangle$ and $\langle Q^3 \cos(3 \delta) \rangle$, were evaluated with the Kumar-Cline sum rules, providing a measure of the electric quadrupole strengths and axial asymmetries. The ${\langle 0_2^+ ||E2||2_1^+ \rangle}$ matrix element was determined for the first time, providing a total, converged measure of the electric quadrupole strength, $\langle Q^2 \rangle$, of the first excited $2_1^+$ level relative to the $0_1^+$ ground state. The results indicate an $E2$ strength that is quenched with respect to both the geometric and U(5) vibrational expectations. The total extracted $E2$ strength is consistent with other theoretical approaches including the shell-, beyond mean-field, and geometric rotor models. However, the shell model did not reproduce the large axial asymmetry found experimentally. These models all appear to have one feature in common: the $2_1^{+}$ state is not vibrational but more similar to a rotation. 

\par%
This manuscript has been authored by UT-Battelle, LLC under Contract No. DE-AC05-00OR22725 with the U.S. Department of Energy. This work was also supported in part by the U.S. DOE under Contract Nos. DE-AC02-06CH11357 (ANL), DE-AC02-05CH11231 (LBNL, GRETINA), DE-AC52-07NA27344 (LLNL, CHICO2), DOE Grant Nos. DE-FG02-97ER41041 (UNC), DE-FG02-97ER41033 (TUNL), and DE-SC0014442 (WU), and by the Australian Research Council under Grant No. DP170101673. This research used resources of ATLAS-ANL, which is a DOE Office of Science User Facility. The publisher acknowledges the US government license to provide public access under the DOE Public Access Plan (http://energy.gov/downloads/doe-public-access-plan).

\bibliographystyle{elsarticle-num}
\bibliography{Cd106}

\end{document}